\documentclass[conference,10pt]{IEEEtran}
\usepackage{verbatim}
\usepackage{cite,authblk,rotating,setspace,amsmath,amssymb,theorem,epstopdf}
\usepackage[caption=false, font=footnotesize]{subfig}

\usepackage{booktabs}
\usepackage{multirow}

\IEEEoverridecommandlockouts
\allowdisplaybreaks

\begin{document}

        \title{Multivariate and Multi-step Traffic Prediction for NextG Networks with SLA Violation Constraints}
        \author[1,2]{Evren Tuna}
        \author[2,3]{Alkan Soysal}
        \affil[1]{\normalsize Department of Research \& Development, ULAK Communications Inc., Ankara 06510, Turkey}
        \affil[2]{\normalsize Department of Electrical and Electronics Engineering, Bahcesehir University, Istanbul 34353, Turkey}
        \affil[3]{\normalsize Bradley Department of Electrical and Computer Engineering, Virginia Tech, Blacksburg, VA 24061, USA}
        \maketitle
\vspace{-.3in}

\begin{abstract}
This paper focuses on predicting downlink (DL) traffic volume in mobile networks while minimizing overprovisioning and meeting a given service-level agreement (SLA) violation rate. We present a multivariate, multi-step, and SLA-driven approach that incorporates 20 different radio access network (RAN) features, a custom feature set based on peak traffic hours, and handover-based clustering to leverage the spatiotemporal effects. In addition, we propose a custom loss function that ensures the SLA violation rate constraint is satisfied while minimizing overprovisioning. We also perform multi-step prediction up to 24 steps ahead and evaluate performance under both single-step and multi-step prediction conditions. Our study makes several contributions, including the analysis of RAN features, the custom feature set design, a custom loss function, and a parametric method to satisfy SLA constraints.
\end{abstract}

\section{Introduction}

Mobile networks are designed to support various vertical applications, each with unique requirements. To efficiently manage these networks, operators must deploy appropriate infrastructure and enhance their predictive capabilities using machine learning (ML) and an SLA-driven orchestration. This allows the networks to adapt dynamically to changing traffic demands. Therefore, it is crucial to develop a well-designed method for predicting cell-based traffic volume, which can be used to manage the allocation of network slices accurately.

The network traffic data in mobile networks depends on space and time due to interactions between base stations (BSs) and daily/weekly trends in mobile data usage. Several studies, including \cite{Bega2020, Zhang2018b, PatrasZhang2018, Fang2018, Zhao2020}, have proposed using Convolutional Neural Network (CNN)-based models to leverage spatial dependencies. Among these studies, \cite{Bega2020, Zhang2018b, PatrasZhang2018} have utilized a 3D-CNN structure borrowed from video processing applications. They assume that the inputs at a given time are in a matrix format, where each entry represents the aggregated traffic of BSs in a corresponding square grid area. On the other hand, \cite{Fang2018, Zhao2020} have proposed using graph convolutional networks (GCNs) instead of a grid structure. In \cite{Fang2018}, the authors have combined Recurrent Neural Networks (RNNs) with GCNs for multi-step prediction. However, their proposed model performs worse than the vanilla LSTM for the special case of single-step prediction and about the same as ARIMA. In addition to GCNs, \cite{Zhao2020} has proposed using handover data to improve performance. The results in terms of Mean Square Error (MSE) and Mean Absolute Error (MAE) reported in \cite{Zhao2020} are 10-15\% better than those of the vanilla LSTM results.

Several studies (e.g., \cite{Wang2017, Chen2018, Feng2018, Alawe2018, Chergui2020, Tuna2022}) have proposed RNN-based models that utilize the temporal dependency of network traffic data. However, in \cite{Alawe2018, Chergui2020}, where the goal is to optimize resource allocation using predicted traffic, evaluating the performance of network traffic prediction is challenging. Meanwhile, \cite{Wang2017, Chen2018} have also utilized a grid structure similar to \cite{Zhang2018b, PatrasZhang2018, Bega2020}, while \cite{Feng2018} has used a private dataset and cell clustering based on the similarity of time-series trends. In a previous work \cite{Tuna2022}, we proposed RNN-based models that used handover data for single-step prediction. However, \cite{Tuna2022} does not guarantee an SLA violation rate.

Most literature uses standard loss functions such as MAE, MSE, and their variants to minimize prediction error \cite{Zhang2018b, PatrasZhang2018, Fang2018, Zhao2020, Wang2017, Chen2018, Feng2018, Alawe2018, Chergui2020}. These symmetrical loss functions treat SLA violation and overprovisioning the same and result in nearly identical SLA violation and overprovisioning rates. However, operators tend to favor overprovisioning to SLA violation due to the strict obligations of SLAs. Nearly 50\% SLA violation resulting from the use of standard loss functions is unacceptable for operators. Therefore, designing and using application-specific and parameterized custom loss functions are crucial.

In this paper, our main goal is to predict DL traffic volume while minimizing overprovisioning and meeting a given SLA violation rate. Our approach is multivariate, multi-step, and SLA-driven, and we make the following contributions:

\begin{itemize}
  \item We analyze 19 different RAN features, in addition to DL traffic volume, to assess their effect on prediction.
  \item We design a custom feature set based on peak traffic hours.
  \item We extract the spatiotemporal effect of high mobility using incoming and outgoing handover relationships between cells.
  \item We perform multi-step prediction up to 24 steps ahead. 
  \item We propose a custom loss function that guarantees the SLA violation rate while minimizing overprovisioning.
  \item We evaluate our approach's performance under single-step and multi-step prediction scenarios.
\end{itemize}

\section{Dataset and Methodology}
\label{section:section_System_Model}

\begin{figure} [t]
	\centering
	\hspace{0.2in}
	\includegraphics[width=0.8\columnwidth]{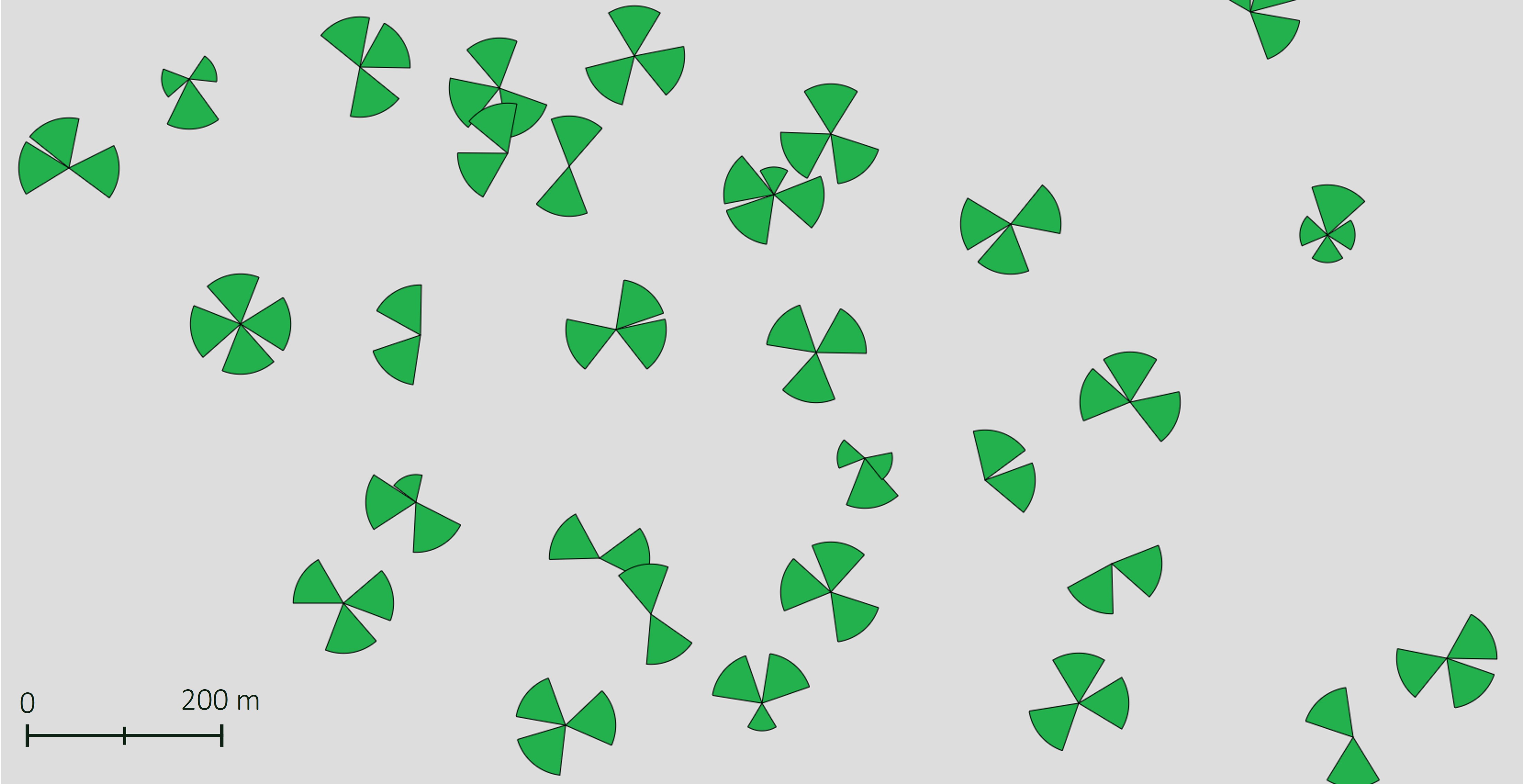}
	\caption{Cells in a metropolitan area.}
	\label{fig:HighlyDenseUrbanArea}
\end{figure}

This section discusses the dataset and methodology used in this study, which focuses on predicting DL traffic volume in a live LTE RAN serving a highly dense urban area with high user mobility. Multiple BSs cover the area, each with varying numbers of sectors and carriers (see Fig.~\ref{fig:HighlyDenseUrbanArea}). Throughout this study, we denote a BS with a two-letter name such as GU, VO, etc. The cells within the relevant BS are represented using two digits, the first being the sector and the second being the carrier number. For example, GU14 refers to the first sector and fourth carrier of the GU base station. Therefore, a carrier in a sector of a BS is referred to as GU14, VO13, etc.

To improve the network's reliability and prevent SLA violations, the goal is to develop an ML-based prediction method that considers unexpected traffic spikes during peak traffic hours. For this purpose, this study uses a multivariate LSTM architecture, which has a higher learning capacity and overcomes the long-term dependency limitation of traditional RNN models.

\begin{figure} [t]
	\centering
	\vspace{0.1in}
	\includegraphics[]{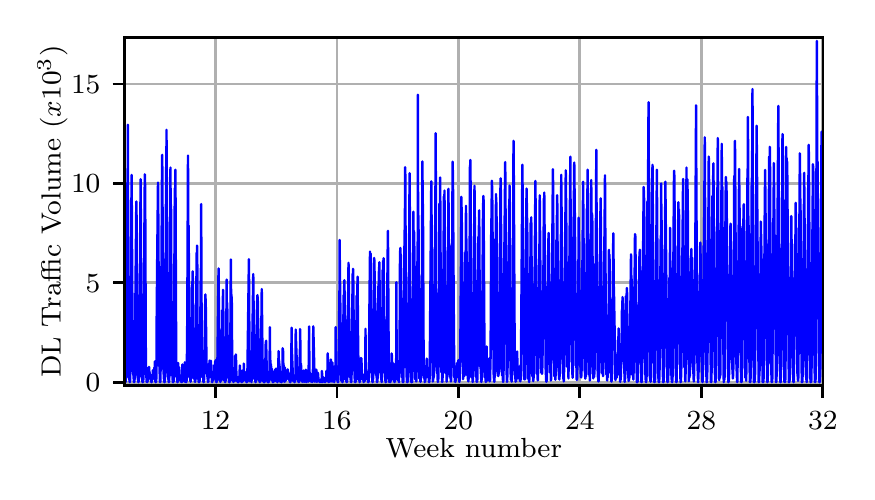}
    \vspace{-15pt}
	\caption{Downlink traffic variation of the GU14 cell.}
	\label{fig:GU14yearlytraffic}
\end{figure}

The dataset contains hourly raw data measurements over 52 weeks. In Fig.~\ref{fig:GU14yearlytraffic}, we show the DL traffic volume graph of the GU14 cell, serving a crowded city square. This cell has a higher traffic load than others, and the trend of the service demand changes throughout the measurement period. Thus, demand spikes occur unexpectedly. 

Table~\ref{table:Features_Abbreviations} lists collected features with their measurement family name \cite{ETSI.132.425, Etsi.132.401}. The dataset used in this study contains 20 RAN features, with DL traffic volume (labeled F10) as the output feature to be predicted. We split the dataset into training, validation, and test sets with lengths of 40 weeks, 8 weeks, and 4 weeks, respectively. We normalize the feature sets using mean and standard deviation values calculated from the training set. Finally, we tune hyperparameters such as learning rate, epoch number, L2 regularization penalty, LSTM or layer number, and hidden unit number using a grid search.

\begin{table} [t]
\centering
\footnotesize
\caption{RAN Features}
\label{table:Features_Abbreviations}
\begin{tabular}{@{}lc@{}}
\toprule
\textbf{Label} & \multicolumn{1}{c}{\textbf{Name}} \\ \midrule
F1  & Num. of Initial E-RABs Attempted to Setup        \\ \hline
F2  & RACH Setup Succ. Rate            \\ \hline
F3  & Avg. RACH Timing Advance                    \\ \hline
F4  & Num. of RRC Attempts                    \\ \hline
F5  & Num. of S1 Signalling Establishment Attempt           \\ \hline
F6  & DL PDCP Cell Thr.      \\ \hline
F7  & UL PDCP Cell Thr.      \\ \hline
F8  & DL PDCP User Thr.      \\ \hline
F9  & UL PDCP User Thr.      \\ \hline
F10 & DL Traffic Volume       \\ \hline
F11 & UL Traffic Volume       \\ \hline
F12 & Avg. UL RSSI Weight PUCCH \\ \hline
F13 & Avg. UL RSRP PUSCH             \\ \hline
F14 & Avg. UL RSRP PUCCH             \\ \hline
F15 & Avg. CQI                         \\ \hline
F16 & Avg. Num. of Active Users in DL           \\ \hline
F17 & Avg. Num. of Active Users in UL           \\ \hline
F18 & Num. of Avg. Simultaneous RRC Connected Users        \\ \hline
F19 & DL PRB Utilisation                    \\ \hline
F20 & UL PRB Utilisation                    \\ \hline
\end{tabular}
\vspace{-12pt}
\end{table}

While training the model, we use the K-fold cross-validation technique to mitigate the changing traffic levels and trends. The number of cross-validation folds, K, is set to six, dividing the 12-month dataset into 2-month intervals. Each fold has different training and validation sets that are obtained by shifting the training dataset by 2 months. 

\section{Parametric Traffic Prediction}
\label{section:section_TrafficPrediction}

The core of our approach is a multivariate and multi-step prediction model for traffic volume, which utilizes an SLA-based weighted loss function. In this section, we will present our proposed weighted loss function, as well as our multivariate feature design and multi-step prediction structure.

\subsection{Weighted Loss Function}

We use $x_t$ to denote the traffic volume measured at time $t$ and $\hat{x}_t$ to denote its corresponding prediction. During test time, we consider SLA violations to occur when the prediction error is negative, i.e., $\hat{x}_t - x_t < 0$, and overprovisioning to occur when the prediction error is positive, i.e., $\hat{x}_t - x_t > 0$. The SLA violation rate is defined as the percentage of instances where an SLA is violated during test time. We also define overprovisioning volume as the average positive prediction error.

Standard loss functions such as MAE and MSE suffer from two significant disadvantages. Firstly, they impose the same penalty for both SLA violations and overprovisioning. Secondly, they cannot be parameterized with an SLA violation rate. Therefore, we propose a custom loss function that overcomes both of these disadvantages of standard loss functions. 

The weighted Mean Absolute Error (wMAE) loss function can be written as
\begin{align}
\mathcal{L}_{\text{wMAE}}(\hat x - x, w) =  \left\{ 
\begin{array}{ll}
	w |\hat x - x|, & \quad \hat x - x \leq 0 \\
	\hat x - x, & \quad  \hat x - x > 0 \\
\end{array}
\right.
\label{eqn:loss-wMAE}
\end{align}
where $w$ is the weight parameter of the function. We aim to minimize the overprovisioning volume while adhering to a constraint on the SLA violation rate by selecting an appropriate $w$. To this end, we explore two different SLA violation rate scenarios, 3\% and 5\%, and employ a univariate LSTM model. For each of these scenarios, we perform a line search to determine the optimal weights, denoted as $w_{3\%}$, and $w_{5\%}$.

\subsection{Multivariate Features}
\label{subsection:subsection_Multivariate}

The aim of our research is to improve the performance of our model by including additional input features beyond the DL traffic volume. Firstly, we investigate the correlation between the DL traffic volume and the 19 other features monitored in the RAN. We expect that including highly correlated features in the input dataset will have a positive impact on the model's performance. Next, we introduce a custom feature set that emphasizes specific time periods during the day or specific days during the week. Finally, to address the spatiotemporal effect, we propose a feature clustering method, which incorporates the incoming and outgoing handover relationships.

To compare the performance of our multivariate model, we use the univariate LSTM model as a baseline. In the univariate model, the input consists of an array of the past 24 hours of DL traffic volume, and the output is the next instance of DL traffic volume. In contrast, in the multivariate model, the input consists of multiple arrays, each containing the past 24 hours of a feature.

\subsubsection{RAN Features}
\label{subsubsection:subsubsection_RANFeaturesOnly}

\begin{figure}[t]
	\centering
	\includegraphics[width=0.9\columnwidth]{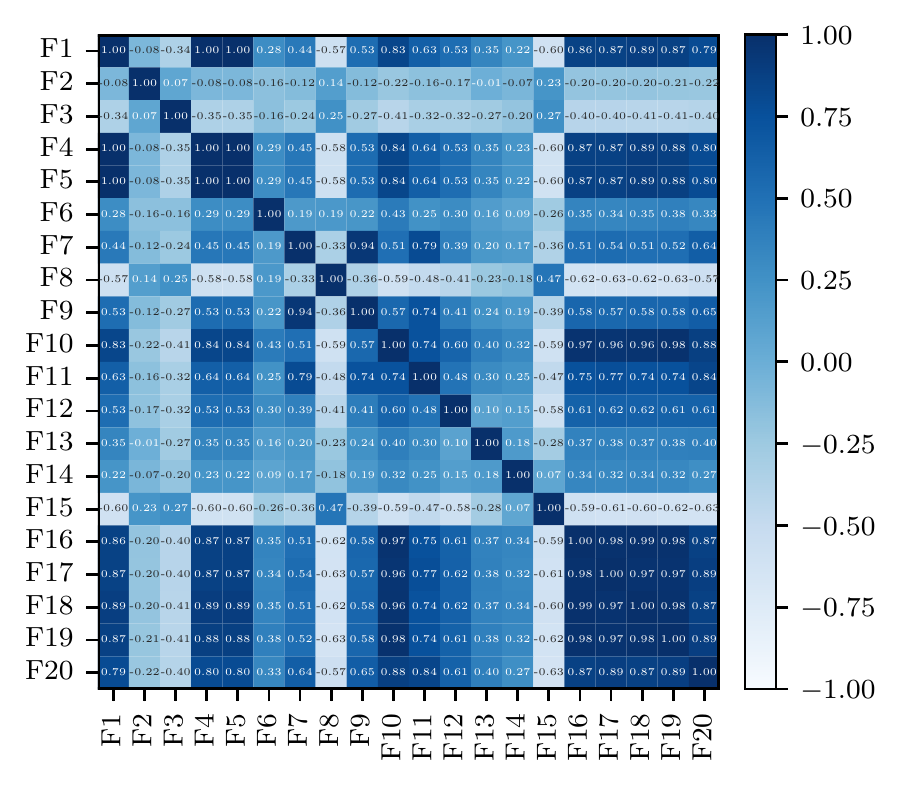}
	\caption{Correlation heatmap of RAN features for GU14 cell.}
	\label{fig:System_Model_Heatmap_All_ENBS}
\end{figure}

We investigate the usefulness of additional input features beyond the DL traffic volume for predicting cell-based multivariate traffic volume. We consider 20 different RAN features, and their labels are presented in Table~\ref{table:Features_Abbreviations}. To identify the most correlated features, we calculate the Pearson correlation coefficient between the DL traffic volume feature (F10) and the other features. Fig.~\ref{fig:System_Model_Heatmap_All_ENBS} shows an example of the correlation heatmap for the GU14 cell. We set a correlation threshold of 0.90 and evaluate the heatmaps of various cells to determine which features can be included in the input dataset. Based on our analysis, we find that the F16, F17, F18, and F19 features are highly correlated with F10 and should be included in the input dataset of our multivariate model. We denote this model as mvLSTM-RAN, which utilizes features F16 through F19 along with F10 to predict the future values of F10. 

\subsubsection{Peak Features}
\label{subsubsection:subsubsection_PeakHours}

We aim to improve our model's performance by addressing the issue of abnormal peaks at busy hours in the F10 feature values. We observe that the F10 feature has a 24-hour period for all cells, with intervals of low and high demand for traffic volume. Moreover, F10 values tend to be higher during weekdays and lower during weekends in densely populated metropolitan areas. To capture these patterns, we introduce a custom Boolean feature vector, ``peak days of the week," to differentiate between weekdays and weekends in the input dataset.

However, our initial model fails to accurately predict F10 values during periods of abnormal peaks at busy hours. To address this, we develop a method to identify peak hours for each cell based on daily 24-hour periods and record the time the F10 value is at its maximum for each period. Using occurrence statistics, we label hours with occurrence values higher than a defined threshold as peak hours. This leads us to propose an additional custom Boolean feature vector, ``peak hours of the day'' to differentiate between peak hours from non-peak hours.

Combining both Boolean features, we introduce our next multivariate model, mvLSTM-peak, which incorporates the peak days of the week and peak
hours of the day vectors in addition to F10. Notably, these additional features are derived from the statistics of F10 and do not require any further RAN measurements.

\subsubsection{Handover Clustering Features}
\label{subsubsection:subsubsection_Handover}

In this section, we investigate the impact of handovers on predicting traffic volume in different cells of the RAN. In densely populated areas with highly mobile users, there are dynamic changes in traffic volume demand as users move between cells.

One approach to exploit the correlation in the handover process is to add the features of all cells in the region to the input dataset to predict the traffic volume feature of the GU14 cell. However, this method significantly increases the problem's complexity and may reduce the model's performance. Another approach is to consider the traffic volume data of only a cluster of cells around the GU14 cell. However, we found that selecting cells in the cluster based only on proximity and/or intersection of coverage does not improve prediction performance, which highlights the limitations of the grid structure of CNN-based approaches. 

\begin{table}[t]
\centering
\footnotesize
\caption{The rates of incoming and outgoing handovers of GU14}
\label{tab:handovergu14}
\begin{tabular}{@{}llll@{}}
\toprule
\multicolumn{2}{c}{\textbf{Incoming}} &
\multicolumn{2}{c}{\textbf{Outgoing}} \\ \cmidrule(r){1-2}  \cmidrule(l){3-4}
  \multicolumn{1}{l}{\textbf{Cells}} &
  \multicolumn{1}{l}{\textbf{Rate \%}} &
  \multicolumn{1}{l}{\textbf{Cells}} &
  \multicolumn{1}{l}{\textbf{Rate \%}} \\ \midrule
    GU12 & 66.79 & GU12 & 26.86 \\
    MS34 & 6.08  & SY24 & 17.24 \\
    VO14 & 5.69  & VO14 & 12.48 \\
    SY24 & 4.68  & GU17 & 8.72  \\
    VO12 & 4.45  & MS34 & 8.31  \\
    GU24 & 3.36  & GU24 & 4.54  \\
    MS37 & 2.05  & GU13 & 3.96  \\
    SY22 & 1.99  & VO12 & 1.88  \\
    GU13 & 1.49  & VO13 & 1.88  \\
    GU22 & 1.15  & RE37 & 1.55  \\ \bottomrule
\end{tabular}
\vspace{-12pt}
\end{table}

We introduce the {\em Handover Clustering} method to exploit the correlation in cell-based incoming and outgoing handover statistics. This method considers only those cells that have a handover relationship with the target cell GU14. Table~\ref{tab:handovergu14} presents the neighboring cells that have incoming and outgoing handover relations with GU14. Our analysis shows that 66.79\% of incoming handover data to GU14 is from the GU12 cell. With Handover Clustering, we expand the input dataset by adding two feature vectors. These vectors are constructed by calculating the weighted averages of the DL traffic volumes of the neighboring cells, one vector for the incoming and the other for the outgoing handover relationship. We refer to our next multivariate model, mvLSTM-handover, which includes the two custom feature vectors formed through the Handover Clustering method.

In summary, we present four multivariate LSTM models for predicting traffic volume in the RAN. The first model, mvLSTM-RAN, incorporates additional RAN features, while the second model, mvLSTM-peak, focuses on the peak hours of the day and days of the week. The third model, mvLSTM-handover, incorporates the weighted averages of DL traffic volumes of cells in the handover cluster. Lastly, we propose the mvLSTM-all model, which combines all three multivariate models.

\subsection{Multi-step Prediction}
\label{subsection:subsection_Multistep}

In this subsection, we examine the performance of our multivariate LSTM models in a multi-step prediction scenario. Specifically, we consider 1-hour, 2-hour, 4-hour, 8-hour, and 24-hour ahead predictions and analyze the performance degradation as we increase the prediction horizon.

In a multi-step prediction scenario, previously predicted values are used as valid past values for subsequent predictions. Therefore, the accuracy of the initial prediction is crucial in determining the performance of the subsequent predictions. To evaluate the performance degradation, we compare the results of the multi-step prediction with those of the single-step prediction.

We perform the multi-step prediction using all four of our proposed models: mvLSTM-RAN, mvLSTM-peak, mvLSTM-handover, and mvLSTM-all. For each model, we conduct experiments on the same test dataset for 1-hour, 2-hour, 4-hour, 8-hour and 24-hour ahead predictions.

\section{Results}
\label{section:section_Results}

In this section, we provide the results of our DL traffic volume prediction for the GU14 cell. Initially, we demonstrate the change in the SLA-based loss and the overprovisioning volume required for 3\% and 5\% SLAs. We then compare the performance outcomes of single-step and multi-step predictions.

\subsection{Different SLA Violation Rate Percentages}
\label{subsection:section_Results_CompSLAs}

\begin{table}[t]
\centering
\footnotesize
\caption{1-hour ahead prediction performance for the GU14 cell}
\label{tab:benchmark-1step-GU14}
\begin{tabular}{@{}lllll@{}}
\toprule
\multirow{2}{*}{\textbf{Models}} &
  \multicolumn{2}{c}{\textbf{3\%}} &
  \multicolumn{2}{c}{\textbf{5\%}} \\ \cmidrule(l){2-3}  \cmidrule(l){4-5}
 &
  \multicolumn{1}{c}{\textbf{Loss}} &
  \multicolumn{1}{c}{\textbf{Volume}} &
  \multicolumn{1}{c}{\textbf{Loss}} &
  \multicolumn{1}{c}{\textbf{Volume}} \\ \midrule
  univariate LSTM & 0.50 & 42.61 & 0.44 & 36.92 \\
  mvLSTM-RAN          & 0.49 & 42.74 & 0.43 & 37.65 \\
  mvLSTM-peak          & 0.46 & 39.23 & 0.42 & 34.75 \\
  mvLSTM-handover          & \textbf{0.44} & \textbf{38.08} & \textbf{0.39} & \textbf{31.28} \\
  mvLSTM-all          & 0.48 & 39.55 & 0.41 & 33.23 \\ \bottomrule
\end{tabular}
\vspace{-12pt}
\end{table}

In this section, we present the performance results of DL traffic volume prediction for the GU14 cell under 3\% and 5\% SLAs. Table~\ref{tab:benchmark-1step-GU14} presents the single-step performance results for the GU14 cell under both SLA conditions.

The results show that the test loss and accompanying overprovisioning volume of the models decrease as the SLA percentage increases. The weight for 3\% SLA is higher than 5\%, causing the loss function to penalize SLA violation cases more severely under the 3\% SLA condition. Consequently, the model avoids violating SLA more, leading to increased test loss and overprovisioning volume due to frequent overestimation of the actual values.

For both SLA conditions, the mvLSTM-handover model performs better than the other models. Specifically, the mvLSTM-handover model yields 12\% and 11.36\% lower test losses than univariate LSTM for 3\% and 5\% SLA conditions, respectively.

\begin{figure} [t]
	\centering
	\includegraphics[width=\columnwidth]{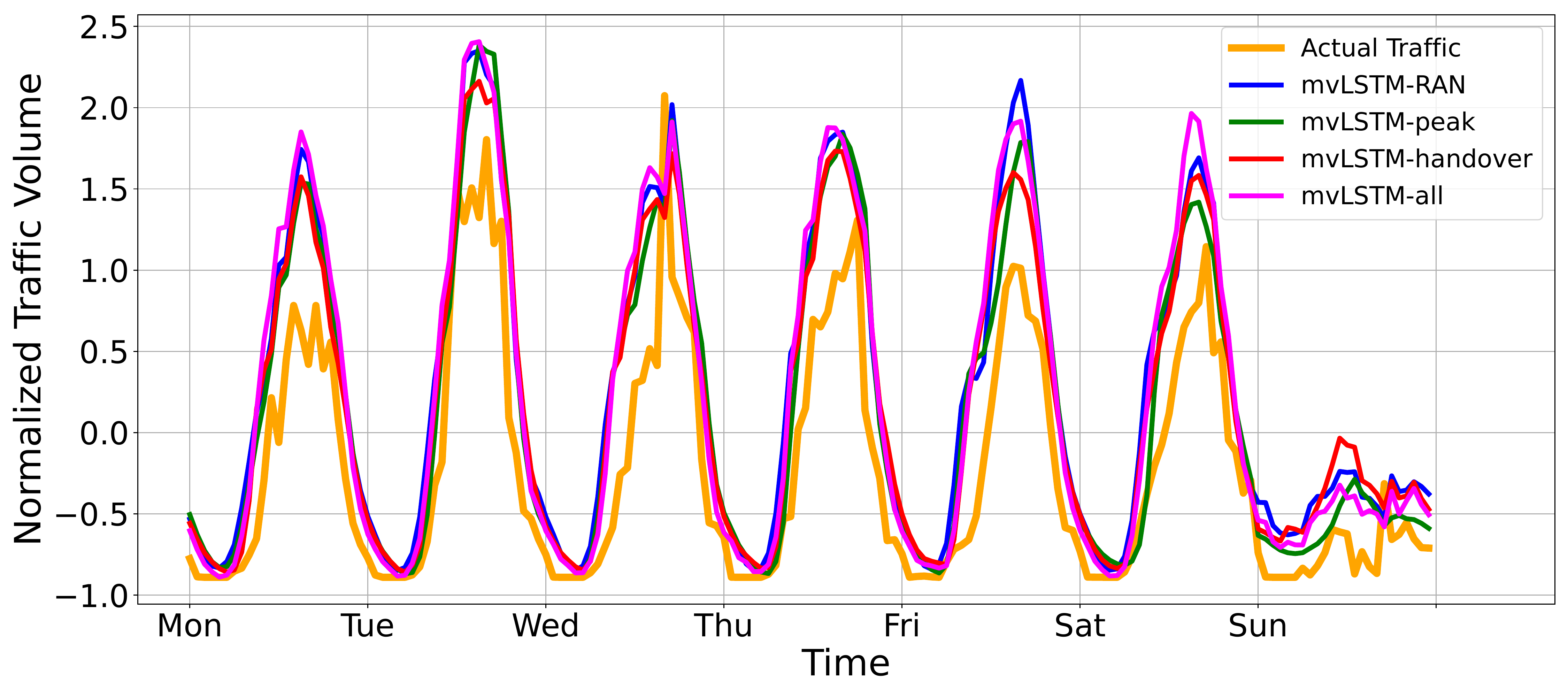}
	\caption{Actual and predicted values for GU14 cell under 3\% SLA.}
	\label{fig:GU14_Diff_SLAs_3P}
\end{figure}

However, the use of additional RAN features in mvLSTM-RAN and mvLSTM-all models resulted in more test loss and overprovisioning compared to mvLSTM-peak and mvLSTM-handover models. Fig.~\ref{fig:GU14_Diff_SLAs_3P} illustrates that mvLSTM-RAN and mvLSTM-all models overestimate more than mvLSTM-peak and mvLSTM-handover models during the late afternoon hours for 3\% SLA. This leads to an increase in the test loss and overprovisioning volume for mvLSTM-RAN and mvLSTM-all models more than mvLSTM-peak and mvLSTM-handover models.

\subsection{Multi-step Prediction}
\label{subsection:section_Results_singlemulti}

\begin{figure}[t]
	\centering
	\includegraphics[width=\columnwidth]{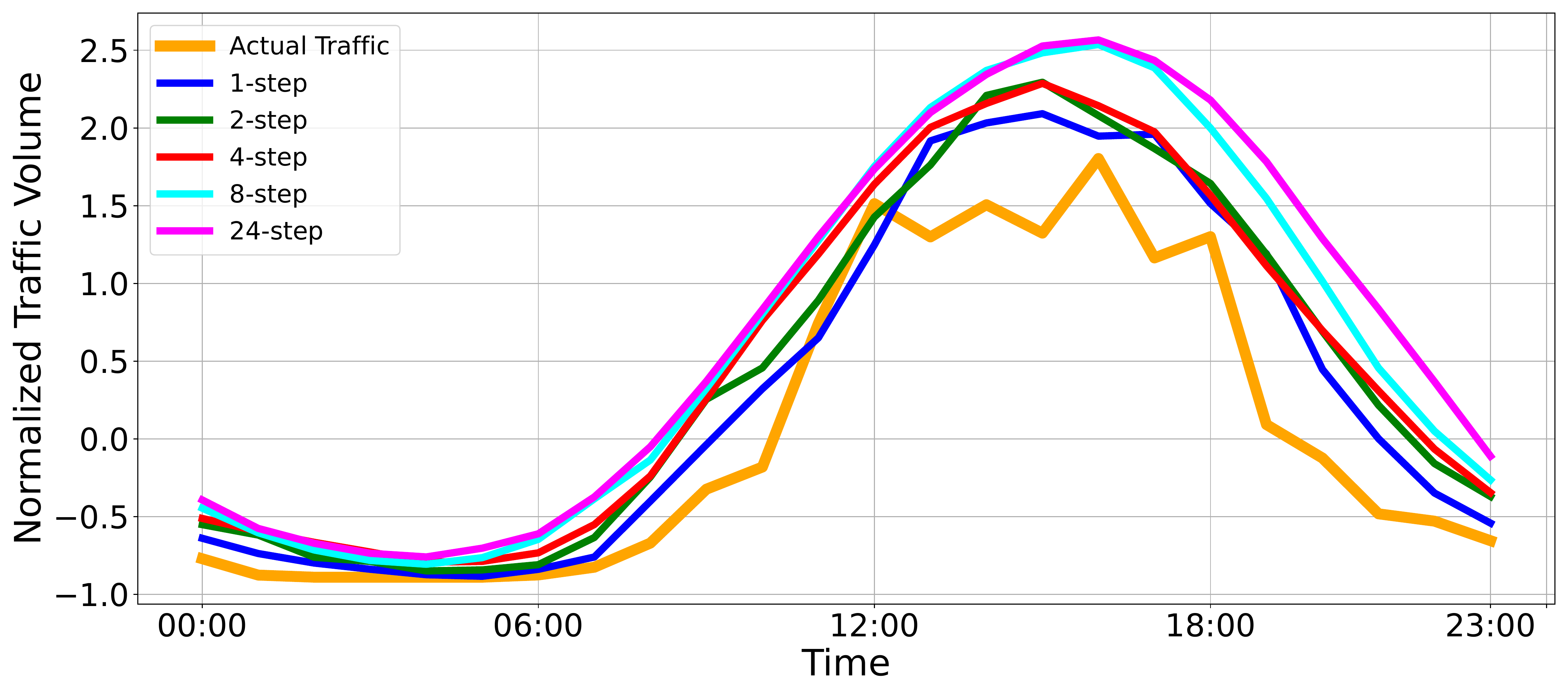}
	\caption{Multi-step prediction using the mvLSTM-handover model.}
	\label{fig:GU14_MS_Time}
\end{figure}

In this section, we present the results of the multi-step prediction for the GU14 cell under the 5\% SLA condition. Table~\ref{tab:benchmark-multistep-GU14} shows the multi-step performance results for the GU14 cell. We observe that the accuracy of the initial prediction significantly affects the subsequent predictions, and the performance degrades as we increase the prediction horizon. As in the 1-hour ahead prediction, the mvLSTM-handover model has the lowest test loss and overprovisioning volume in the 2-hour ahead prediction. As the number of prediction steps increases, the performance of all models degrades, resulting in SLA violation when traffic demand increases and overprovisioning when it decreases. The test loss for the mvLSTM-handover model's 2-hour ahead prediction is 18\% higher than that of the 1-hour ahead prediction. In multi-step prediction, increasing the number of steps leads to the model doing more overprovisioning during periods of increasing and decreasing traffic demand. Specifically, during the evening hours, when the traffic demand begins to decrease, we observe a lag of up to 3 hours in 24-hour predictions. The time graph in Fig.~\ref{fig:GU14_MS_Time} illustrates the delay in the model predictions as the number of steps increases.

\section{Conclusion}
\label{section:section_Conclusion}

In this paper, we present four multivariate LSTM models for predicting DL traffic volume for a particular cell. We compare the performance of these models under two SLA conditions, 3\% and 5\%, for both single-step and multi-step predictions.

Our results show that the mvLSTM-handover model outperformed the other models under both SLA conditions for single-step prediction. On the other hand, the use of additional RAN features in mvLSTM-RAN and mvLSTM-all models do not provide better performance compared to mvLSTM-peak and mvLSTM-handover models.

In the multi-step prediction scenario, the performance of all models degraded as the number of steps increased. Similar to the 1-hour ahead prediction, we observe that the mvLSTM-handover model yields the lowest test loss value compared to other models in the 2-hour ahead prediction. However, when the number of steps is 4 or more, only considering the handover relationship makes it difficult to capture the changes in the neighbor cells of GU14.

In conclusion, our study demonstrates the effectiveness of multivariate LSTM models for DL traffic volume prediction in cellular networks. Furthermore, our findings suggest that incorporating handover clustering with weighted averages of DL traffic volumes of neighboring cells can significantly improve the prediction performance under different SLA conditions. The study recommends that operators set thresholds for degradation compared to single-step prediction and adjust their time horizon accordingly.
\begin{table}[t]
\centering
\footnotesize
\caption{Multi-step prediction test loss results for GU14 cell}
\label{tab:benchmark-multistep-GU14}
\begin{tabular}{@{}lccccc@{}}
\toprule
\textbf{Models} & {\textbf{1-hour}} & {\textbf{2-hour}} & {\textbf{4-hour}} & {\textbf{8-hour}} & {\textbf{24-hour}}
\\ \midrule
  univariate LSTM   & 0.44          & 0.61          & 0.70          & 0.87          & 0.91 \\
  mvLSTM-RAN        & 0.43          & 0.49          & 0.56          & 0.77          & 1.00 \\
  mvLSTM-peak       & 0.42          & 0.66          & \textbf{0.48} & \textbf{0.65} & \textbf{0.82} \\
  mvLSTM-handover   & \textbf{0.39} & \textbf{0.46} & 0.57          & 0.89          & 1.07 \\
  mvLSTM-all        & 0.41          & 0.54          & 0.59          & 0.69          & 0.95 \\ \bottomrule
\end{tabular}
\vspace{-12pt}
\end{table}

\end{document}